\documentclass[conference]{IEEEtran}
\usepackage{graphicx}

\begin{document}
\title{Performance Evaluation in High-Speed Networks by the Example of Intrusion Detection}

\author{
    \IEEEauthorblockN{Thomas Lukaseder, Jessika Fiedler, Frank Kargl}
	\IEEEauthorblockA{Institute of Distributed Systems\\
	Ulm University, Germany\\
	\{firstname\}.\{lastname\}@uni-ulm.de}
}

\maketitle

\begin{abstract}
Purchase decisions for devices in high-throughput networks as well as scientific evaluations of algorithms and technologies need to be based in measurements and clear procedures. Therefore, evaluation of network devices and their performance in high-throughput networks is an important part of research.
In this paper, we document our approach and show its applicability for our purpose in an evaluation of two of the most well-known and common open source intrusion detection systems, Snort and Suricata.
We used a hardware network testing setup to ensure a realistic environment and documented our testing approach.
In our work, we focus on accuracy of the detection especially dependent on bandwidth. We would like to pass on our experiences and considerations.
\end{abstract}
\begin{IEEEkeywords}
IDS; performance tests; high-speed networks 
\end{IEEEkeywords}

\section{Introduction}
\label{chp:intro}

Computer networks provide a wide range of targets for malicious purpose. An attacker can be able to steal sensitive data or even bring the system to a standstill. For this reason, there must be mechanisms to protect networks. Intrusion detection systems (IDS) are one of the most important components of a modern security infrastructure. Especially communities and companies that deal with a large and growing amount of data have increasing demands on current computer networks. This is the reason why 10~Gbps Ethernet has become a standard within company and university networks. In terms of security, IDS are commonly used to inspect networks and as an essential component to find different kinds of attacks. As they observe network traffic, network-based systems are now facing a great challenge and it must be tested, whether they can handle it. There is a clear need for an overview of the different systems and their performance in high-speed network traffic, so provider get an idea of what solution to choose. Such an overview is challenging as the term Intrusion Detection System is used for a wide area of technologies. Therefore it must be discussed how those different solutions can be evaluated in a comparative way. As the results are supposed to be comparable, it must be ensured that all of the solutions run under exactly the same conditions. This means, they should get the same resources in terms of hardware and input data. Based on our previous work in building network testing frameworks~\cite{bradatsch1}\cite{bradatsch2} and testing networks~\cite{lukaseder}, we evaluate IDS in high-throughput networks. The paper is organised as follows: Section~\ref{chp:related} lists comparable works our work is based on, Section~\ref{chp:ids} elaborates on the IDS we chose to evaluate and why while Section~\ref{chp:evaluation} documents our approach. In Section~\ref{chp:results} the most important evaluation results are summarized. Section~\ref{chp:conclusion} concludes the paper.

\section{Related Work}
\label{chp:related}

There is a lot of work discussing IDS and their ability to handle high-speed traffic. Most of them focus on improving the settings of software and systems to the new requirements. When it comes to performance evaluations in high-speed networks, there are only a couple of detailed reports available.
Concerning the general evaluation of IDS, Milenkoski et al.~\cite{Milenkoski2015} should be mentioned. The authors give a detailed overview of practices, that are used to evaluate IDS. They structure the evaluation into three parts: workloads, metrics and methodology. We use this as a guideline for this project and help to generate benchmark data. In the last part they present measurement methodologies, which define the systems properties of interest as well as the employed workloads and metrics to evaluate this properties. However, they do not conduct any tests.
The three open-source solutions of Intrusion Detection Systems\,---\,namely Suricata, Bro and Snort\,---\,have been tested extensively. Those tests mostly focus on their accuracy. Only a few of them evaluate their performance in high-speed networks. Khalil~\cite{Khalil2015} gives a good overview of the three open-source solutions. He outlines their problems with handling high-speed traffic and the solutions those systems implement. He also outlines a couple of performance tests that have been run on those systems. They analyzed overall performance concerning traffic throughput but did not analyze the precision of the IDS under test.
Bul\'ajoul et al.~\cite{Bul2013} ran a couple of tests to evaluate Snorts performance. For their analysis they altered three different values: the number of packets per second, number of packets sent over all and the packet length. As performance indicators they use the number of packets received, analyzed, dropped, and filtered. The number of packets filtered does not occur in the figures, only the number of packets sent. According to the authors this value is not altered during the experiment, only the speed at which those packets have been sent. There is a figure included where the number of packets sent decreases when the speed increases. This is contrary to the point that the number of packets sent is not altered. Furthermore, this value should be redundant to the number of packets received as Snort should receive the packets sent, however, the data indicates otherwise. Nevertheless, their experiments show that the number of packets dropped increases, when they increase the speed. This could give a basic indicator of Snorts performance in high-speed traffic.
Most IDS answer the challenge to handle high-speed networks with some kind of load-balancing. Vallentin et al.~\cite{Vallentin2007} present a cluster solution for network-based intrusion detection. They introduce Cluster Bro, which is a solution to handle high-speed traffic with Bro. The input is split across several 'worker' nodes, that analyze their part of the traffic and a management system is used to perform the overall analysis. Their work also includes a performance evaluation to show that the cluster produces sound results even at high throughput. They also mention that their solution ran at IEEE Supercomputing 2006, where it 'monitored the conference’s primary 1 Gbps backbone network as well as portions of the 100 Gbps High Speed Bandwidth Challenge network which is a very broad term and could mean anything between 0 and 100 Gbps. The basic points are that the cluster gives the same results as a single Intrusion Detection System. In their performance evaluation they focus on the ability to balance the incoming traffic to the different nodes\,---\,namely the scalability\,---\,and the overhead of communication this cluster solution introduces in comparison to a single IDS. There is no evaluation of the Cluster's accuracy as part of this paper.

\section{Systems under Review}
\label{chp:ids}

There are three relevant open-source IDS that can be found in related work\,---\,Bro, Suricata, and Snort. We limit our analysis to Snort and Suricata as Bro is an anomaly-based IDS that functions differently to the other two in many ways.
Snort\footnote{snort.org} is an open-source network-based Intrusion Detection System. Its functionalities include packet capturing, analysis of captured packets and\,---\,important for this work\,---\,live analysis of network traffic. The version tested in this paper is 2.9.9.
A lot of dependencies have to be installed before Snort is able to run properly. Those include libnet\,---\,a network API to gain access to different protocols, libpcap\,---\,a library to capture network traffic and pcre\,---\,a collection of functions with Perl semantic and syntax. New to version 2.9 is the usage of the DAQ library, which replaces libpcap calls to simplify packet-I/O-options.\\
Snort provides the possibility for developers and users to add own modules called preprocessors. One of them is Perfmon~\cite{Perfmon}. Using this preprocessor the user gets informed about statistics such as received and dropped packets during runtime. To use the preprocessor a corresponding option has to be set during the build process.

After Snort has been installed successfully, it needs to be configured. This is done using the configuration file written in a specific Snort format. For this work, configuration meant to set network variables, choose active rules, activate the Perfmon preprocessor, and set output formats. There are three different sets of rules: subscriber, registered, and community rules. The basic set of rules is taken from the subscriber snapshot and is extended by rules to detect included attacks.

Another open-source network-based Intrusion Detection System is Suricata\footnote{suricata-ids.org}.
Its version 3.2.1 is the basis for testing. Besides the same dependencies as Snort, it needs libraries to process files in yaml format. This is used to read and write the configuration file. Different from Snort is the multi-threading option. This is included in all Suricata versions and needs to be configured correctly. Its performance is depending on the number of processors and some memory caps that have to be set properly with regards to the chosen number of processors. Fine tuning of Suricata is done according to~\cite{Regit2012}. During a first experiment, it became obvious that the number of processors influences the performance, and with more processors memory consumption grows extremely.

Suricata is able to parse and use Snort rules. In addition, there is a set of rules called emerging rules~\cite{SuricataRules}, which has been used as a basis. During live analysis, Suricata produces alert logs in the same format as Snort. Furthermore, Suricata outputs statistics on the current values of received and dropped packets.

\section{Evaluation Setup}
\label{chp:evaluation}

Two networks have to be simulated: an external network and a home network with the IDS in the middle. Two server are used to simulate the networks. One of them to send malicious and benign packets and the other to receive them. Both are connected via 10 GB interfaces to the IDS server. 
The server has to have sufficient resources to be capable of analyzing the traffic at line speed. We chose an off-the-shelf server with 4 CPU cores with 3.1 GHz\footnote{IntelXeon Processor E3-1220 v3} and 6 GB of memory; multi-threaded intrusion detection systems can make use of 4 threads per processing step.

\subsection{Traffic Simulation}
\label{sec:pakets}
\begin{figure*}[h]
	\centering
	\includegraphics[width=0.7\textwidth]{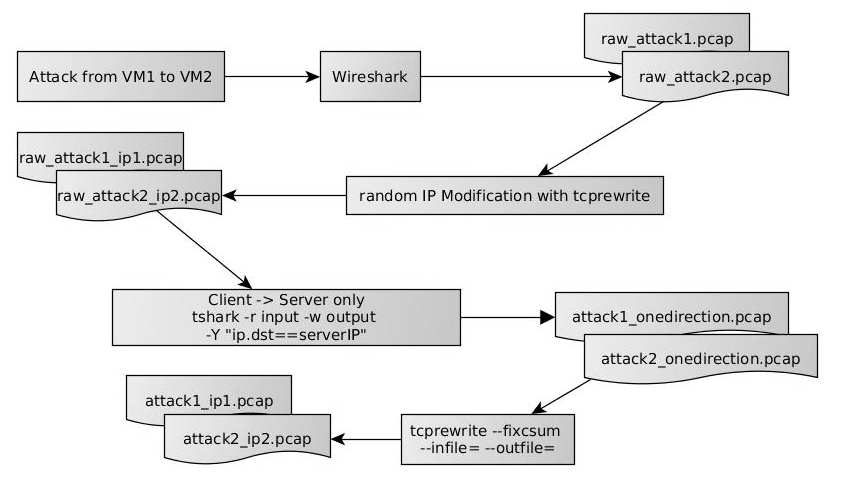}
	\caption{Generating malicious traffic captures.}
	\label{fig:paketflow}
\end{figure*}
Malicious traffic is simulated and captured before the actual evaluation so repeatability of the tests is ensured. The attacking VM is running Kali Linux, whereas the target machines run Ubuntu Server, one with Metasploit. Wireshark is used to capture traffic. Target IP addresses are set to the receivers incoming interface. The attacking IP address is changed in a manner that each attack has its own sending IP address. This is done to distinguish logged alerts. As the response of the receiving server is not analyzed by the IDS under test, they are removed from the captured traces. Figure~\ref{fig:paketflow} shows the complete process of generating attack captures.

We used the threat report by McAfee Labs\cite{mcafee} to determine a typical attack mix that realistically represents the day to day threats of networks connected to the Internet. According to the report, browser-based attacks are the most prevalent, followed by brute force login attempts (e.g. SSH brute force), followed by denial of service attacks, SSL attacks, scans (e.g. nmap scans), DNS spoofing, and backdoors. All in all, these attacks account for 91\% of all network attacks. We generated these attacks with different tools. The whole list can be seen in Table~\ref{tab:attacks}. 
\begin{table}[b]
    \centering
	\begin{tabular}{l|l|}
		Attack type & Tool used \\
		\hline
		\hline
		successful SSH brute force & Metasploit framework\\
		unsuccessful SSH brute force & Metasploit framework\\
		TCP connect flood & nping\\
		TCP SYN flood & hping3\\
		UDP flood & hping3\\
		SYN scan & nmap -sS\\
		SYN OS-scan & nmap -sS -O\\
		UDP scan & nmap -sU\\
		User enumeration & nmap
	\end{tabular}
	\caption{Included attack types.}
	\label{tab:attacks}   
\end{table}
The benign traffic is simulated with iperf3\footnote{iperf.org}, as it is capable of generating 10 Gbps traffic at runtime mixed with data peaks.
To simulate a data center it is necessary to include data peaks. Large datasets are publicly available at different sources (KDnuggets~\cite{kdnuggets} as an example). Those captures can later be resent several times per evaluation run. A problem is to gain appropriate traffic speed with tcpreplay, this could only be achieved when storing the data set in a RAM-disk.

\subsection{Rule Selection}
\label{sec:rules}

Appropriate rules for the IDS must be chosen to ensure that they are able to detect all attacks included in the attack traffic\,---\,given that the IDS has the time to analyze the traffic. Furthermore, the traffic has to be labeled to later decide if a detection is a false or true positive. The rule sets for the IDS under test follow different syntax but have to be semantically identical to ensure that speed differences between the IDS only depend on the IDS implementation. For the performance evaluation a minimum set of rules has been chosen as the purpose is to measure the capability to handle high-speed traffic not the influence of the number of rules. Thresholds for detection and similar parameters are chosen identical for both systems (e.g. 150 as the threshold for flooding attack alerts).
The number of active rules influences the overall performance. Therefore, this number is the same for both IDS.

\subsection{Evaluation Process}
\label{sec:test}
A single test needs four parameters: the time for the evaluation, the number of attacks per minute, the traffic speed, and the chosen IDS. During one test, three phases can be distinguished: initialization, evaluation, and output. During the first step the IDS is initialized and the attack plans for the setup are read from configuration files. Those files contain a list of attacks that should be send at minute $m$ during an evaluation of $M$ minutes. This ensures that each IDS is tested with the same attacks and that there are no variations within a test run.
During the test run, traffic, CPU, and memory usage is monitored. Benign traffic is sent continuously at the given speed and the defined attacks per minute using tcpreplay. In this setup, more attacks  also results in a slightly higher number of packets per minute. Table \ref{tab:progSteps} shows the steps on each server during a single test.
\begin{table*}[t]
    \centering
	\begin{tabular}{l|l|l|l|}
		Program part & IDS & Sender & Receiver\\
		\hline
		\hline
		Start & start IDS	& read in attacks & \\
		\hline
		& \multicolumn{3}{|c|}{wait for IDS}\\
		\hline
		Monitoring & CPU, memory, traffic & outgoing traffic & incoming traffic\\
		\hline
		Evaluation part & wait & send packets & wait \\
		\hline
		Output & \multicolumn{3}{|c|}{modify output files}\\
		\hline
		End & kill IDS instance & & \\		
		\hline
		& \multicolumn{3}{|c|}{Resting phase}\\
		\hline
	\end{tabular}
	\caption{Per server tasks during the evaluation program.}
	\label{tab:progSteps}
\end{table*}
The output of a certain throughput and different amount of attacks is one test sample (e.g. 5,10,15,20 attacks at 2~Gbps). A test phase includes all test samples for a chosen IDS. This means tests are performed for throughputs and attack amounts in given ranges. A sample test and a complete test phase are both completely automated. The processes on all three servers must be started at the same time, synchronized clocks on all servers are required.

\subsection{Result Processing}
\label{sec:output}
Each test has several outputs in different formats. The result processing for a single test\,---\,meaning exactly one value for speed and one for attacks per minute\,---\,are discussed first. Timestamps must be compared to ensure that all outputs are within the same time period. Most traffic speed log files differ in their units. They are recomputed to Gb and the different bandwidth files are combined to one. This can be used to check the resulting traffic throughput during evaluation and find bottlenecks if those exists. As mentioned earlier, all included IDS output information about their current received and dropped packets. Those values are extracted and combined with the measured CPU and memory usage. The CPU documentation gives the amount of CPU usage per core.
Due to some rule settings, it is possible that a reference log file includes messages that are not necessary to identify an attack. This especially occurred when detecting SSH brute force attacks with Suricata. During the reference phase, logs contain messages such as 'ET INFO NetSSH SSH Version String Hardcoded in Metasploit'. Those are legitimate messages but not necessary to identify the actually attack\,---\,however, the message 'ET SCAN Potential SSH Scan' is. To keep track of messages that can appear but do not have to, priority files have been introduced. They include the attack message and a priority\,---\,0 means the message must be there and 1 it is acceptable if it is missing. The value is only added to false positives if the message itself really was not expected in this minute. Some messages are redundant. As an example Snort either logs a 'TCPFilteredScan' or 'TCPScan' message when detecting a SYN scan. Those messages are mapped so they end up as correctly logged alerts. The result of the process is a file containing a mapping of minute, message to logged- and expected counters as well as a statistic counted per type of attack. The first file can be used to identify the (mis-)matches per minute during one test. Furthermore, those values are used to identify the detection performance. 
Once all tests of a test sample have been processed, sample wide values can be counted. This means per minute attack detection statistics are summed up and used to compute performance measurements. One of them is true positive rate or sensitivity which represents the detected attacks out of all attacks. Another value of interest is precision giving the percentage of correctly identified alerts among all logged alerts.
CPU and memory consumption are averaged over all measured values. Packet information is taken from the IDS performance output and from system log files. Suricata outputs totals of the current values, Snort averages over runtime, therefore values are computed into averages over runtime for all packet information files. The same has to be done for alerts per second values and the drop rate. 
The number of packets that have not been considered can be computed from the number of packets that have been send and the number of packets that have been analyzed by the IDS. All output values for a test sample are summarized in Table~\ref{tab:perSample}.\\
\begin{table*}[t]
    \centering
	\begin{tabular}{l|l|l|l|}
		Value & Meaning & Arithmetic\\
		\hline
		\hline
		True Positive TP & Correct logged messages & sample sum \\
		\hline
		False Positive FP & Logged but not expected & sample sum \\
		\hline
		False Negative FN & Expected but not logged & sample sum \\
		\hline
		\hline
		True Positive Rate & Attack detection rate (Sensitivity) & TP/(TP + FN) \\
		\hline
		False alarm rate & Rate of false alarms & FP/(TP + FP)\\
		\hline
		Precision & Rate of correct alerts among all alerts & TP/(TP + FP) \\
		\hline
		\hline
		CPU & CPU usage of IDS & sample average \\
		\hline
		Memory & Memory usage of IDS & sample average  \\
		\hline
		Received packets RP & Packets analyzed by IDS & average over time \\
		\hline
		Droprate DP & Packets dropped by the IDS & average over time \\
		\hline
		Send packets SP & Actual send packets & average over time \\
	\end{tabular}
	\caption{Result values per test sample.}
	\label{tab:perSample}
\end{table*}

\section{Results}
\label{chp:results}

\begin{figure}
\centering
\includegraphics[width=0.85\columnwidth]{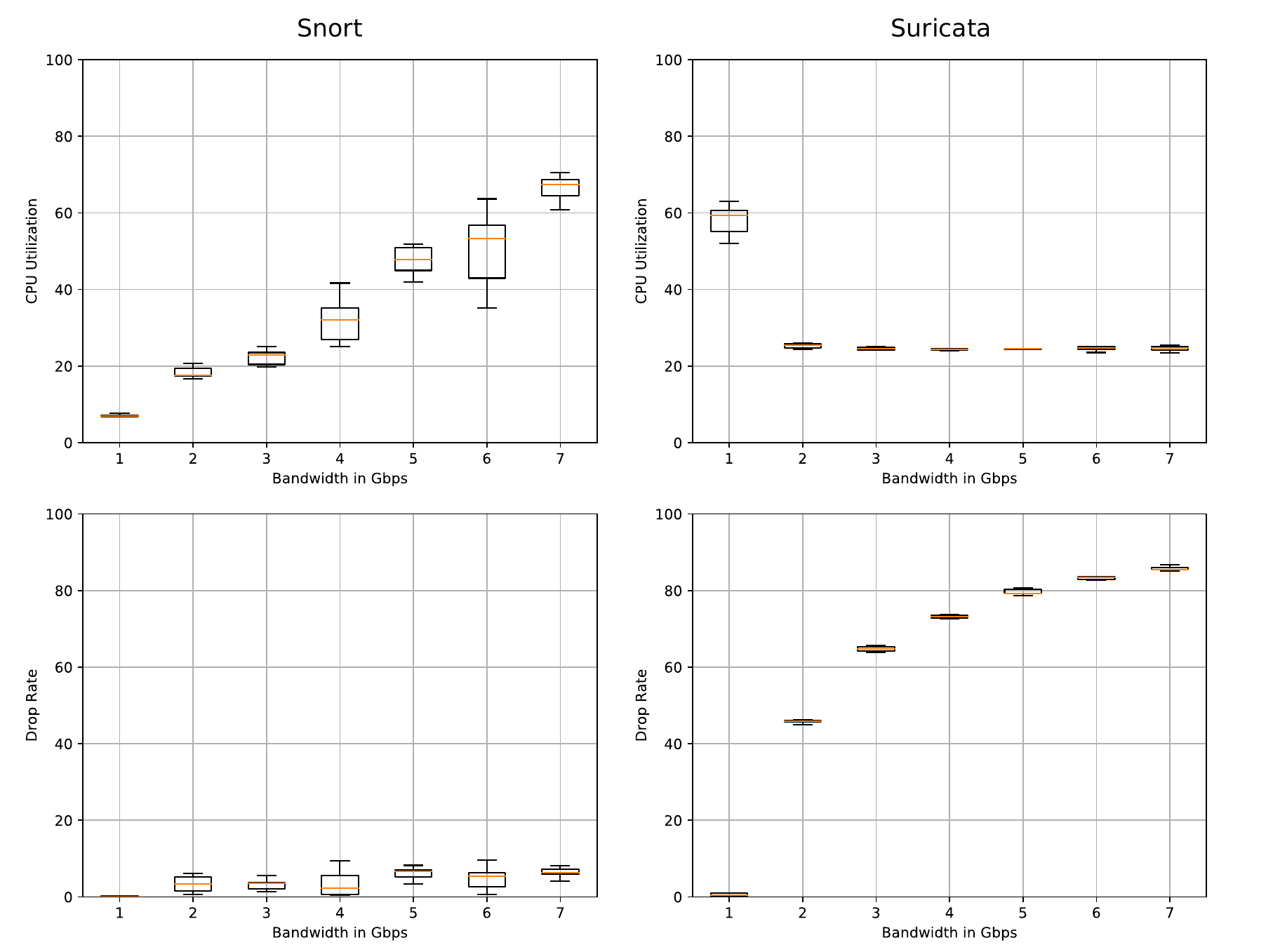}
\caption{CPU utilization per CPU core in use and packet drop rate of Snort and Suricata dependent on bandwidth.}
\label{fig:results-bandwidth}
\end{figure}

\begin{figure}
\centering
\includegraphics[width=0.85\columnwidth]{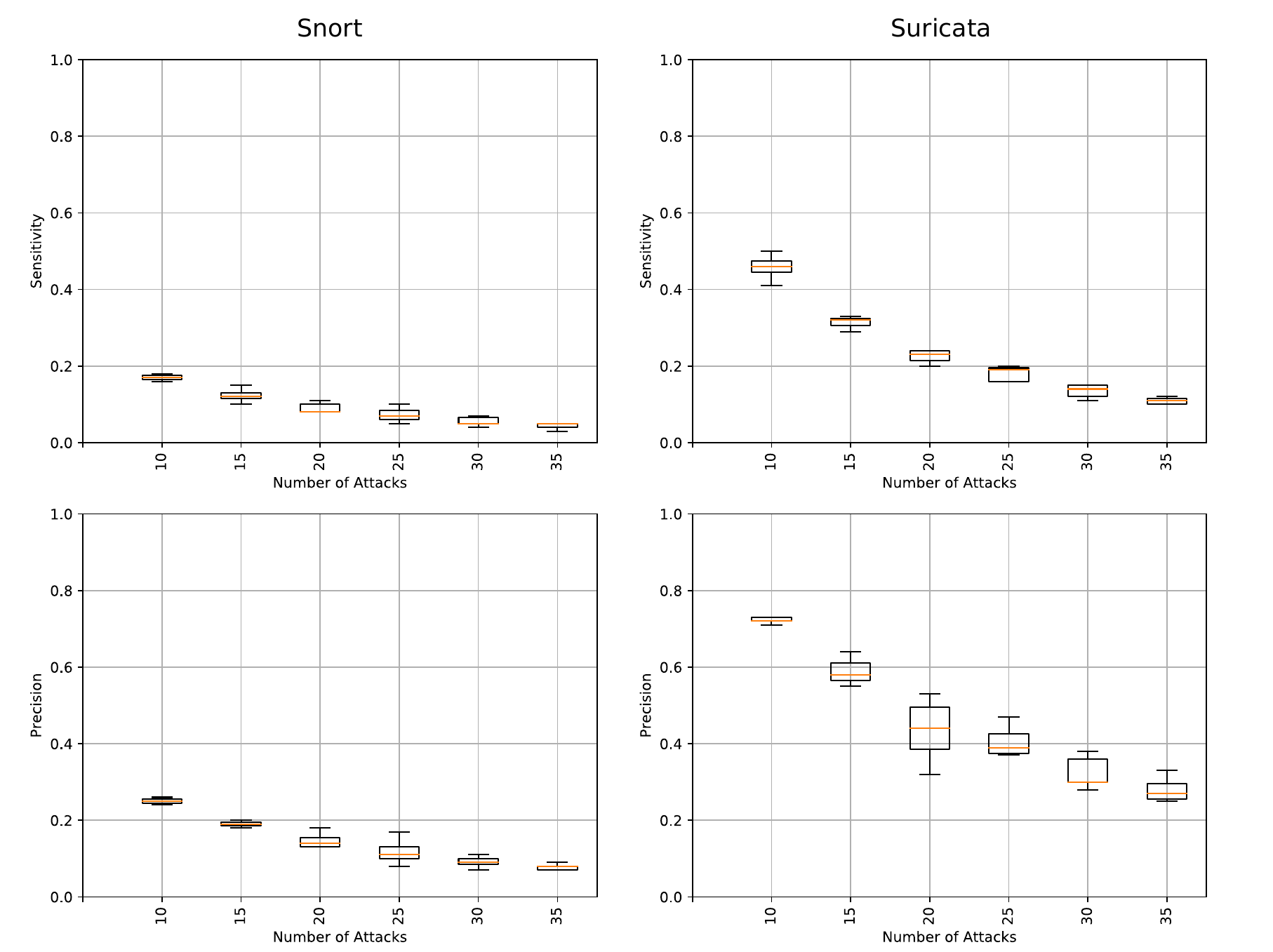}
\caption{Precision and sensitivity dependent on the amount of attacks per minute.}
\label{fig:results-attackers}
\end{figure}

The evaluation part included tests for 30 minutes each with equal attack distributions. Meaning all types of attacks mentioned previously are included with the same probability. Traffic speeds from 1 to 7 Gbps have been included as well as attacks from 10 to 35 per minute.
The data implicates, that throughput has great influence on the CPU utilization and drop rate while the amount of attacks does not influence these factors. Precision and sensitivity on the other hand where not influenced by the throughput (despite the high drop rate differences) but only by the amount of attacks in the network. Neither bandwidth nor amount of attackers had influence on memory usage (Snort: 6MB, Suricata: 80MB). There were no false positives observed in any test scenario.

Figure~\ref{fig:results-bandwidth} shows the CPU and packet drop measurements. Every data point contains measurements with different amount of attackers (10, 15, 20, 25, 30, and 35 attackers per minute). While Snort's drop rate is low throughout all measurements, the CPU utilization rises linearly and reaches up to 65\% with 7 Gbps of traffic. Suricata's approach is different. The CPU utilization per CPU core stays quite constant other than the measurement with only 1 Gbps of traffic as less CPU cores are in use here.
 However, Suricata begins to drop packets as soon as the throughput is higher than 1 Gbps. Suricata consistently analyses 1 Gbps and drops any traffic exceeding that limit.

Figure~\ref{fig:results-attackers} shows the precision and sensitivity measurements. Every data point contains measurements with different throughputs of attack traffic (1, 2, 3, 4, 5, 6, and 7 Gbps). Although Suricata heavily drops packets depending on throughput, and Snort does not, higher throughput did not influence the precision or sensitivity. However, more attacks in the test traffic did lead to worse detection rates. All in all, Suricata showed better performance.

\section{Conclusion and Future Work}
\label{chp:conclusion}

In this work, we described a possible way to handle performance evaluation tasks and used this process to evaluate two software-based IDS and stress tested their performance. We used a hardware-based network setup with commodity hardware, tested with different throughput and different amounts of attacks per minute to evaluate which parameters influence performance the most and what administrators have to keep in mind when choosing and setting up an IDS in their network. We showed that bandwidth only plays a minor role for the precision of the systems under review, while the amount of expected attack traffic should be considered. Furthermore, the systems under review tend to heavily rely on good CPU performance while their memory requirements are comparably low.

For the future, we will extend our evaluation to other IDS (e.g. Bro) and to test beyond 10~Gbps on stronger hardware to evaluate whether software IDS can be applicable in backbone networks. We plan to update the data set regularly based on the threat reports and\,---\,together with the configuration files for the IDS\,---\,publish this data. Furthermore, we plan to extend the evaluation work to include other security devices such as firewalls.

\section*{Acknowledgment}

This work was supported in the bwNET100G+ project
by the Ministry of Science, Research and the Arts Baden-
W\"urttemberg (MWK). The authors alone are responsible for
the content of this paper.

\bibliographystyle{IEEEtranS}
\bibliography{bibliography}
%
\end{document}